\documentclass[fleqn,usenatbib]{mnras}

\usepackage{booktabs,caption}
\usepackage[flushleft]{threeparttable}

\usepackage[T1]{fontenc}

\DeclareRobustCommand{\VAN}[3]{#2}
\let\VANthebibliography\thebibliography
\def\thebibliography{\DeclareRobustCommand{\VAN}[3]{##3}\VANthebibliography}

\usepackage{graphicx}	
\usepackage{amsmath}	
\usepackage{amssymb}	
\usepackage{multirow}
\usepackage{newtxtext,newtxmath}

\title[The GC distribution in the NGC 1052 group]{The large-scale structure of globular clusters in the NGC~1052 group}

\author[M. L. Buzzo et al.]{
Maria Luisa Buzzo$^{1,2}$\thanks{E-mail: lgomesbuzzo@swin.edu.au},
Duncan A. Forbes$^{1,2}$,
Jean P. Brodie$^{1,2,3}$,
Steven R. Janssens$^{1,2}$, 
Warrick J. Couch$^{1,2}$, 
\newauthor
Aaron J. Romanowsky$^{3,4}$, 
and Jonah S. Gannon$^{1,2}$
\\ \\
$^{1}$ Centre for Astrophysics and Supercomputing, Swinburne University, John Street, Hawthorn VIC 3122, Australia \\
$^{2}$ ARC Centre of Excellence for All Sky Astrophysics in 3 Dimensions (ASTRO 3D), Australia \\
$^{3}$ Department of Astronomy \& Astrophysics, University of California Santa Cruz, 1156 High Street, Santa Cruz, CA 95064, USA \\
$^{4}$ Department of Physics and Astronomy, San José State University, One Washington Square, San Jose, CA 95192, USA \\
}

\date{Accepted 2023 March 27. Received 2023 March 27; in original form 2022 December 23}

\pubyear{2022}

\begin{document}
\label{firstpage}
\pagerange{\pageref{firstpage}--\pageref{lastpage}}
\maketitle

\begin{abstract}
Prompted by the many controversial claims involving the NGC~1052 group, including that it hosts two dark matter-free galaxies with overluminous and monochromatic globular cluster (GC) systems, here we map out the large-scale structure (LSS) of GCs over the entire group.
To recover the LSS, we use archival optical CFHT imaging data.
We recover two GC density maps, one based on universal photometric properties of GCs from simple stellar population models, and one based on the properties of spectroscopically confirmed GCs in DF2 and DF4 (the two dwarf galaxies with overluminous GC populations). 
Both selection methods reveal overdensities around the massive galaxies in the group, as well as around NGC~1052 itself, that are coincident with the positions of previously identified stellar streams and tidal features.
No intragroup GCs are found connecting these structures to any of the dwarf galaxies. We find, however, two other dwarfs in the group hosting GC systems. These include RCP32 with 2 GCs with ages equivalent to the GCs around NGC~1052, and DF9 with 3 GCs with ages similar to the GCs around DF2 and DF4. We conclude that the GC distribution in the group does not strongly support any formation scenario in particular. It favours, nonetheless, scenarios relying on galaxy-galaxy interactions and on the coeval formation of GCs around the DM-free dwarf galaxies. These may include the recently proposed bullet-dwarf formation, as well as high-redshift tidal dwarf galaxy models.
\end{abstract}

\begin{keywords}
globular clusters: general, galaxies: dwarf, galaxies: formation
\end{keywords}



\section{Introduction}
\label{sec:introduction}


The NGC~1052 group has been the topic of intense debate over several years due to claims that it hosts dark matter (DM)-free low-surface brightness (LSB) dwarf galaxies \citep{vanDokkum_18,vanDokkum_19}. This came as a surprise since both dwarf \citep{Strigari_08a,Strigari_08b} and LSB galaxies \citep{deBlok_97} are believed to be some of the most DM-dominated galaxies in the universe. This is because their gravitational potentials are too weak to counteract the outward pressures of stellar feedback, resulting in decreased star formation efficiencies. 

Two galaxies claimed to be largely DM-free within their stellar components, NGC~1052-DF2 and NGC~1052-DF4 (DF2 and DF4, hereafter), were also found to have extremely extended sizes (effective radius, $R_{\rm e}$ > 1.5 kpc) for their surface brightnesses, being better classified as ultra-diffuse galaxies (UDGs). Additionally, they were found to host a population of ultra-luminous globular clusters (GCs) \citep{vanDokkum_18, vanDokkum_19, Shen_21}. All of these peculiar properties are accompanied by the fact that the galaxies are very close in projection, but far away from each other in line-of-sight tip of the red giant branch (TRGB) distance \citep[2 Mpc, ][]{Shen_21}. 
The puzzle created by these complications has culminated in scepticism about the DM-free nature of these galaxies \citep{Trujillo_19, Montes_20}  and have also led to the development of simulations and theoretical methods to understand the possible formation pathways of such galaxies.
Assuming that it is not a coincidence that the galaxies share such unusual properties while being so close in proximity, a common formation scenario for them must simultaneously explain: 1) their lack of DM, 2) their large sizes and 3) the presence of overluminous GCs. 

One of the most commonly proposed scenarios involves tidal stripping by a more massive galaxy 
\citep{Ogiya_18, Maccio_21, Jackson_21, Ogiya_21, Moreno_22}. This scenario potentially explains the DM depletion \citep{Haslbauer_19} and extended sizes of the galaxies, and it is expected to leave behind a trail of stripped GCs, but so far it is not capable of explaining the overluminous GC population around DF2 and DF4.
Alternatively, \cite{Trujillo-Gomez_21}, attempting to explain both the low DM content and the peculiar GC systems, proposed that the galaxies were formed by a combination of an early, intense burst of star-formation that creates a rich GC system with a top-heavy GC luminosity function (GCLF), and feedback that expands the galaxy. The expansion leads to an LSB galaxy residing in a diffuse DM `core' creating the illusion of a DM-free galaxy. While it is an interesting possibility, this model also implies that all GC-rich LSB galaxies or ultra-diffuse galaxies should have unusual GCLFs and be DM-poor -- in contradiction to the observations \citep[e.g.,][]{Toloba_18,Saifollahi_22,Gannon_22a, Gannon_22b}.

Recently, \cite{Silk_19}, \cite{Shin_20} and \cite{Lee_21} have proposed that DM-free galaxies with overluminous GCs could form in a `mini bullet cluster' event \citep{Clowe_06}, where a high-speed collision in the host group would be able to separate baryonic- and dark-matter. These studies were the first ones capable of explaining aspects of the DM depletion and GC population, although not the spatial distribution of gas and GCs.
\cite{vanDokkum_22}, building on this scenario, suggested that not only DF2 and DF4 were formed in the aftermath of this high-speed interaction, but also seven to nine other dwarf galaxies in the NGC~1052 group. The collision would have happened $\sim$8 Gyr ago, separating the baryonic and DM content of the progenitor galaxies. As a result, the gas would have formed a trail of DM-free galaxies (forming a near-linear distribution in projection) along with many massive GCs, while the DM itself would lie at the ends of the trail in two DM-dominated galaxies \citep[see ][figure 1]{vanDokkum_22}. 
This scenario can potentially explain all of the unusual properties of these galaxies, but it makes strong predictions. One of the most important predictions is that all of the galaxies in the trail (except the DM-dominated ones), as well as any formed GCs, should have the same ages and metallicities --hence, colours-- since they would have been formed by the same process, from the same material and at the same time. 

One way of testing this hypothesis is by studying the stellar populations of all of the galaxies in the supposed trail, and checking to see if they have equivalent properties. So far, this has only been done for DF2 and DF4 by \cite{Buzzo_22b}, where they concluded that these share similar stellar populations and are both consistent with an age of $\sim$8 Gyr. Additionally,  \cite{Fensch_19} has measured both the stellar content and GC population of DF2 with VLT/MUSE, finding that they are old and metal-poor, with an age of 8.9 $\pm$ 1.5 Gyr for the stars and 8.9 $\pm$ 1.8 Gyr for the GCs. These measurements are consistent with the findings of \cite{Buzzo_22b} and confirm that the ages (and thus colours) of the GCs and stellar body of DF2 are equivalent.
\cite{vanDokkum_22b} compared the GC systems of DF2 and DF4, showing that they have the same optical colours. Finally, the NGC~1052 group was also studied by \cite{Roman_21}, who found a prominent GC system around one of the galaxies in the trail, RCP32, as well as many LSB galaxies in the group. These LSB galaxies were the input to the methodology applied by \cite{vanDokkum_22} to find this near-linear trail distribution of galaxies in the group.

While all of these observational constraints were obtained for DF2 and DF4, there is so far no stellar population evidence that any of the other galaxies were formed by the same process. That is mainly because all of the other galaxies in the trail are much fainter and, thus, obtaining reliable stellar populations for them is more challenging (note, however, Gannon et al. in prep. for a detailed study of the stellar populations of NGC 1052-DF9 and Tang et al. in prep. for a study of the stellar populations of all galaxies in the trail). 

In this study, we take an alternative path to studying the stellar populations of the dwarfs themselves. We aim to test different formation scenarios for the galaxies in the NGC~1052 group by studying its GC large scale structure (LSS) out to the virial radius of the central galaxy \citep[$R_{\rm vir}$ = 390 kpc,][]{Forbes_19}. We will look for possible imprints that each scenario would have left in the group and see if the recovered distribution is compatible with their predictions. These include the bullet-dwarf, tidal dwarf galaxies (TDGs) and tidal stripping scenarios. For some of the scenarios, imprints are expected not only in the GC systems of the galaxies, but also in the form of intragroup GCs. 
The GC LSS will reveal any GC structures connecting the galaxies, if other galaxies in the group host GC systems and if these GCs share similar stellar populations amongst themselves, elucidating if and how the formation histories of the galaxies in the group are connected.


Although many studies to date have characterised the GC systems of galaxies, few have recovered the LSS of GCs in groups and clusters \citep[e.g.; ][]{Lee_10,Durrell_14,Taylor_17,Madrid_18,Ragusa_22, Chies-Santos_22,Pan_22}. We note here the works of \cite{Lee_10} and \cite{Ragusa_22}, which recovered the whole GC LSS of the Virgo cluster (16.5 Mpc) and Leo I pair (10.8 Mpc), respectively. These were carried out using solely photometric properties (e.g., colours and magnitudes) and have revealed remarkable structural features in the cluster/pair. In a similar approach, we provide in this work the GC LSS of the controversial NGC~1052 group. While this study will be primarily used to evaluate the degree of compatibility of specific formation scenarios with the recovered LSS, it also sheds light on processes and interactions taking place in group environments.

Throughout this paper we assume a distance modulus to the NGC~1052 group of $(m-M)_0$ = 31.5 and distance of 20 Mpc \citep{Shen_21}. {When discussing the bullet-dwarf scenario, we assume that there are eleven candidate galaxies in the proposed trail \citep[see Table 1 of][]{vanDokkum_22}, namely RCP32, NGC~1052-DF2 (DF2), RCP28, RCP26, RCP21, NGC~1052-DF9 (DF9)\footnote{DF9 was not considered part of the trail by \cite{vanDokkum_22}, as it was excluded from the \cite{Roman_21} selection of LSB galaxies in the NGC~1052 group for having a nucleus. As the galaxy has similar visual properties and lies along the same near-linear distribution, we consider it to be part of the trail.}, RCP17, TA21-12000, NGC~1052-DF4 (DF4), NGC~1052-DF5 (DF5), NGC~1052-DF7 (DF7)\footnote{DF7 does not have CFHT coverage.}.

\section{Data}
\label{sec:data}
We used archival Elixir-reduced non-stacked Canada-France-Hawaii Telescope (CFHT) MegaCam data for our study. 
The CFHT data were obtained on 2020 September 12 in the $u$, $g$ and $i$ bands (Program ID: 20BO44). 
To correct for astrometric differences between the exposures, we used \texttt{SCAMP} \citep[version 1.7.0,][]{Bertin_02}. 

The data were combined using \texttt{SWARP} \citep[version 2.15,][]{Bertin_02}. For this, we used twelve images with 990 seconds of exposure time each in the $u$ band, five with 335 seconds in $g$ and five with exposures of 455 seconds in $i$. 
This resulted in final images with exposure times of 11880, 1675 and 2275 seconds in the $u$, $g$ and $i$ bands, respectively. 
The overall seeing of the images is 0.96, 0.80 and 0.76 arcsec, and the magnitude zero points of the $u$, $g$ and $i$ bands are 24.67, 26.26 and 25.51, respectively. We note that Tang et al. (in prep.) found a systematic magnitude difference of $\sim$0.1 mag for every band when comparing stars in the CFHT images and SDSS. This  suggests that there may be extra uncertainties involved when using CFHT data.
Where relevant, flux was converted into magnitudes using the formula provided by CFHT that takes into account the magnitude zero point along with exposure times, airmasses and a colour calibration parameter. All magnitudes and colours are in the AB system.

\section{Analysis}
\label{sec:analysis}
\subsection{\texttt{SExtractor}}

For the source detection, we used \texttt{SExtractor} \citep[version 2.19.5,][]{Bertin_02} in single-mode in each band. GCs at the distance of the NGC~1052 group (20 Mpc) are point sources in the CFHT images. Thus, we defined a threshold of 2$\sigma$ above the background to detect sources, assuming, given our seeing, that point sources should have most of their light within apertures of $\sim$3 pixels diameter (pixel scale of CFHT is 0.206 arcsec per pixel). 
For the aperture photometry, we used a smaller aperture with a diameter of 2 pixels to avoid contamination from the sky or nearby sources in the GC candidate photometry. Then we performed local sky subtraction and added in an aperture correction calculated using the growth curves of isolated bright point sources. The corrections are 0.94, 0.55, and 0.54 mag in the $u$, $g$ and $i$ bands, respectively. Errors in the photometry are the nominal errors from \texttt{SExtractor}.
As the last step, we corrected the data for Galactic extinction using the correction derived by \cite{Calzetti_00} and the dust maps of \cite{Schlafly_11}.

\subsection{GC selection}
\label{sec:GC}

As mentioned before, GCs at a distance of 20 Mpc are point sources in our imaging and, thus, they cannot be distinguished from foreground stars or high-redshift galaxies. To account for that, we have applied both morphological (ellipticity and concentration) and photometric (colours and magnitudes) GC selection criteria. 
The morphological selection criteria are summarised as:
\begin{itemize}
\item Concentration ($C_{3-6}$):\\ Given our seeing, we assume that point sources should have most of their light within 3 pixels. Thus, the magnitude difference measured in 3 and 6 pixels diameter apertures should be consistent with zero, allowing a scatter of 0.1 mag (typical photometric error). 

\item Ellipticity ($\epsilon$): \\Point sources should be nearly circular with $\epsilon < 0.2$. 
\end{itemize}

In addition to these criteria, we adopted two different approaches for photometric selection. The first selection was based on photometric properties of GCs in general and the second on photometric properties of spectroscopically confirmed GCs around DF2 and DF4. \\
For the first approach, we used the universal GCLF and simple stellar population (SSP) models consistent with universal GC populations to select GCs.
The SSP models were constructed using the Flexible Stellar Population Synthesis package \citep[FSPS;][version 0.4.2]{Conroy_09,Conroy_10a,Conroy_10b}. We used the Padova isochrones \citep{Marigo_07,Marigo_08} and allowed for metallicities in the range [$Z$/H] = $-$2.2 to 0.0 dex and ages (t$_{\rm age})$ = 8--14 Gyr.
The predicted CFHT colours of these SSP models are shown in Figure \ref{fig:GC_selection}, where each coloured line corresponds to one metallicity, while the extent of the lines is defined by the age range of the SSPs.

\begin{figure}
    \centering
    \includegraphics[width=\columnwidth]{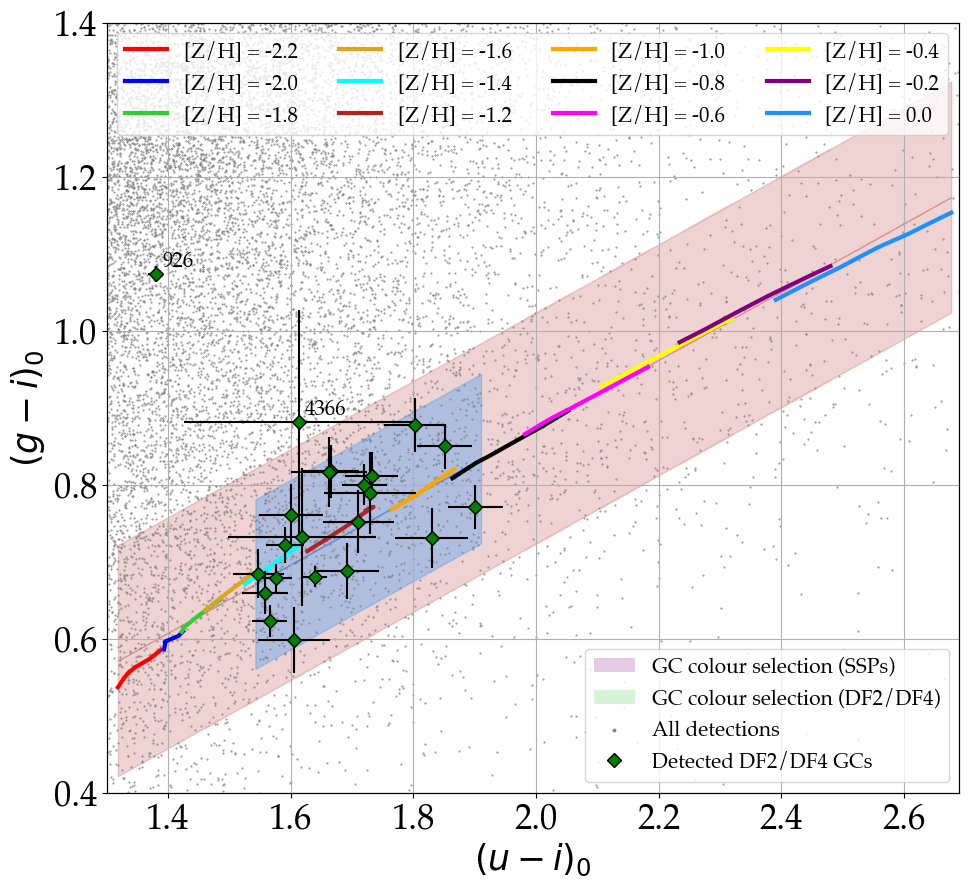}
    \caption{GC selection based on optical colours. Grey dots in the background are all of the detections from \texttt{SExtractor}. Detected DF2 and DF4 spectroscopically confirmed GCs are shown in dark green. The coloured lines show FSPS SSPs with metallicities $-2.0 < $ [$Z$/H] $< 0$ dex and ages $8.0 < t_{\rm age} < 14$ Gyr, and the red-shaded area defines the adopted scatter around the SSPs. GCs selected based on the colours of DF2/DF4 GCs are those within the blue region. GCs \#926 and \#4366 are overlapped by stars in our imaging and were excluded from the selection. All data are corrected for Galactic extinction.}
    \label{fig:GC_selection}
\end{figure}

\begin{itemize}

\item Magnitude: \\
All sources 3$\sigma$ brighter and 1$\sigma$ ($1\sigma = 1.0$ mag) fainter than the peak of the universal GCLF \citep[i.e., $M_g = -7.2$ or $g = 24.3$ mag, ][]{Jordan_07} were selected. The unbalanced cuts (i.e., 3$\sigma$ up and 1$\sigma$ down) were applied based on photometric errors, i.e., only objects with $g$-band photometric uncertainties smaller than 0.2 were selected.

\item Colour ($u-i$ and $g-i$): \\ 
The colour cuts were determined by fitting a line to the SSP distribution in the colour--colour space, allowing a scatter of 3$\sigma$ (0.15 mag). The fit is best parametrized by: 
    \begin{equation}
        (g-i) = 0.44 \times (u-i) - 0.01
        \label{eq:ssp}
    \end{equation}
    
    The selected GCs are those within the red-shaded area in Fig. \ref{fig:GC_selection}.  

\end{itemize}

The second selection consisted of selecting sources with colours and magnitudes similar to those of spectroscopically confirmed GCs around DF2 and DF4. 

\begin{itemize}

\item Magnitude: \\
All sources within $\pm$1 mag of the mean of the GCLF of DF2 and DF4 recovered by \cite{vanDokkum_18}, i.e., $M_g = -8.8$ ($g = 22.7$) mag were selected. 
We note the caveat that DF2 and DF4 were found to have both a population of overluminous GCs and a population of ``normal'' GCs, i.e., consistent with the universal GCLF \citep{vanDokkum_22b}, and by applying this criterion we are including only the overluminous ones.

\item Colour ($u-i$ and $g-i$):\\ 
Colour cuts were defined using the locus of spectroscopically confirmed GCs in DF2 and DF4 in the colour--colour plane (see Fig. \ref{fig:GC_selection}). We assume that these follow the same distribution as the SSP models, but with narrower ranges both in $u-i$ and $g-i$ colours, i.e., $1.54 < u-i < 1.90$ and a scatter of 0.1 mag in $g-i$ . 
\end{itemize}

\begin{figure*}
    \centering
    \includegraphics[width=\textwidth]{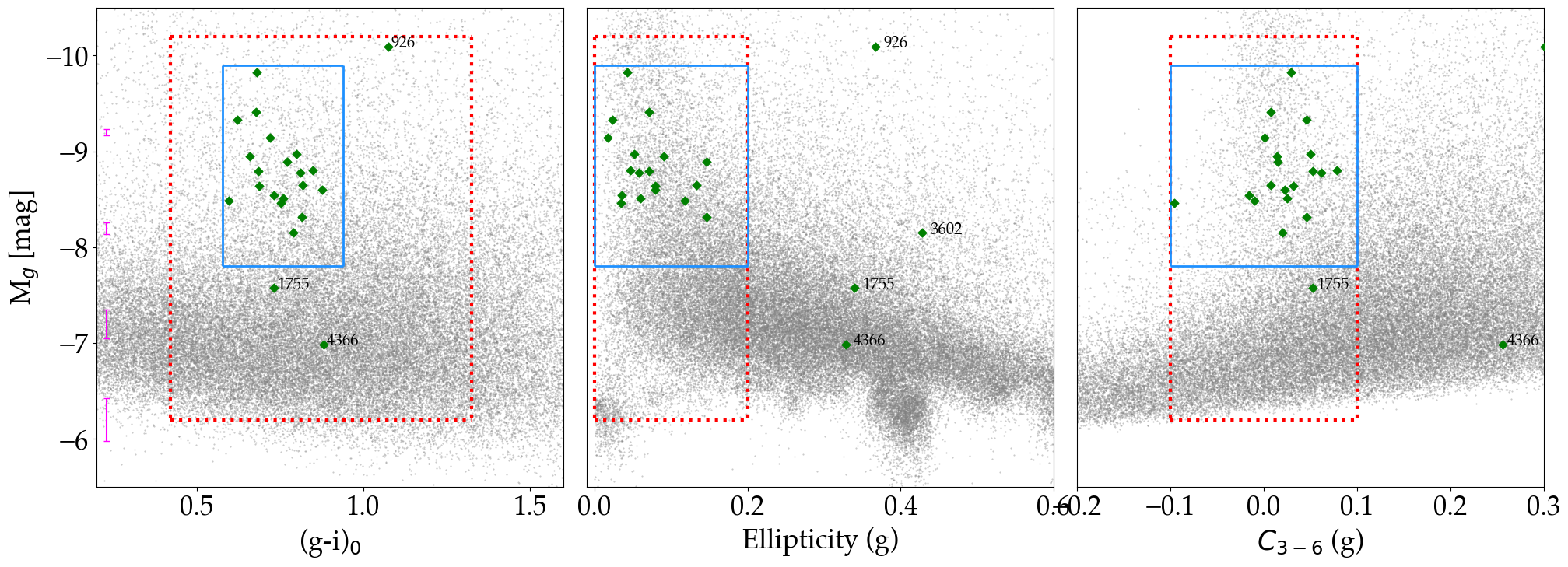}
    \caption{GC selection criteria based on morphological and photometric criteria. In all panels, the detections are shown with the grey circles, and spectroscopically confirmed GCs around DF2/DF4 are shown with the green diamonds. Red-dotted lines show the GC selection based on the properties of GCs in general. Blue lines show the GC selection based on the properties of spectroscopically confirmed GCs around DF2/DF4. \textit{Left:} Color-magnitude diagram. Magenta errorbars show the typical photometric uncertainties at each corresponding magnitude. \textit{Middle:} Ellipticity vs. absolute magnitude. \textit{Right:} Concentration parameter vs. absolute magnitude. GCs \#1755, \#4366, \#3602 and \#926 (not shown in the figure) were excluded from our sample for not meeting the ellipticity, concentration and/or colour criteria.}
    \label{fig:selection_criteria}
\end{figure*}

\begin{table}
    \centering
    \caption{Selection criteria applied to the GC candidates.}
    \begin{tabular}{c|c|c} \hline
    \textbf{Criterion} & \textbf{All GCs} & \textbf{DF2/DF4 GCs colours} \\ \hline
        Concentration ($C_{3-6}$) & $0.0 \pm 0.1$ & $0.0 \pm 0.1$  \\
        Ellipticity ($\epsilon$) & $<0.2$ & $<0.2$ \\
        Magnitude ($M_g$) & --10.2 $< M_g < -6.2$ &  $-9.8 < M_g < -7.8$\\
        $u-i$ & $1.30 < u-i < 2.65$ & $1.55 < u-i < 1.90$ \\
        $g-i$ & Eq. (\ref{eq:ssp}) $\pm \, 0.15$ & Eq. (\ref{eq:ssp}) $\pm \, 0.1$ \\ \hline
    \end{tabular}
    \label{tab:selection}
\end{table}

We note that the spectroscopically confirmed GCs \#926 and \#4366 were excluded from our sample for having different colours to the remaining GCs around DF2 and DF4. Visual inspection has confirmed that these GCs are contaminated by foreground stars in our imaging. GCs \#3602 and \#1755 were also excluded for not meeting the ellipticity criterion (see middle panel of Fig. \ref{fig:selection_criteria}). Thus, in the end of the selection process, we have 20 out of the 24 spectroscopically confirmed GCs around DF2/DF4.

In summary, we selected two samples of GC candidates, i.e., one based on SSP criteria to resemble typical GCs and one based on the spectroscopically confirmed GCs around DF2/DF4. All of the selection criteria are summarised in Table \ref{tab:selection} and shown in Fig. \ref{fig:selection_criteria}.

\subsection{Artificial star tests}
\label{sec:artstars}

To quantify our completeness, we have carried out tests using artificial stars.
We first modelled our PSF using the routine \texttt{DAOPHOT II} \citep{Stetson_87} in the $g$ band, using a sample of 15 isolated, bright and unsaturated stars.
We then injected 30000 artificial stars in the range of $19 < g < 26$ magnitudes onto blank areas of our $g$-band image.
The completeness was calculated by comparing the total number of artificial stars injected in different magnitude bins and the number of detected artificial stars using the same methodology as the one we used to detect the GCs (i.e., \texttt{SExtractor}). 

We find that we are 94\% complete down to the universal GCLF turnover magnitude, i.e., $M_g = -7.2$ ($g$ = 24.3) mag, and a 50\% completeness limit of $M_g = -6.8$ ($g = 25.0$) mag.
We inspected different areas of the images and have found little-to-no difference in completeness throughout our FoV, with the exception of the centre of NGC~1052.

\section{Results}
\label{sec:results}

\subsection{The GC systems of DF2 and DF4}
\label{sec:DF2/DF4}

The spectroscopically confirmed GCs around DF2 and DF4 were used to select a range of new GC candidates throughout the NGC~1052 group. It is important, none the less, to understand how our recovered photometric properties align with other studies in the literature of the GCs around DF2 and DF4.
Upon cross-matching our GC catalogues with \cite{Shen_21}, we detect all 24 spectroscopically confirmed GCs around these galaxies. Out of the 24, only 20 met all of our selection criteria (GCs \#926, \#3602, \#1755 and \#4366 were excluded) and are the ones discussed in what follows.

The GCLF reconstructed from these 20 GCs around DF2 and DF4 reveals a mean value at $M_g = -8.8$ ($g = 22.7$), consistent with the findings of a top-heavy GCLF for DF2 and DF4 by \cite{vanDokkum_18,vanDokkum_19} and \cite{Shen_21}. We investigated the extent that the age of a GC can influence its brightness to understand if this top-heavy GCLF could be explained by an age difference other than mass. The magnitude difference of a GC of age 8 Gyr and a GC of 14 Gyr at fixed stellar mass and metallicity is 0.5 mag in the $g$ band. This difference is less than a third of the magnitude difference observed between the peak of the universal GCLF and the one found for DF2 and DF4. We thus conclude that the overluminous nature of the GCs is unlikely due to age and most likely due to mass.
The locus of the overluminous GCs around DF2/DF4 in the colour--colour plane is consistent with the SSPs with metallicities $-1.4<$[$Z$/H]$<-1.0$ dex, which is in agreement with the findings of \cite{Fensch_19} and \cite{Buzzo_22b}.
They have an average colour of $\langle g-i \rangle = 0.76 \pm 0.10$ and $\langle u-i \rangle = 1.66 \pm 0.12$. These values correspond to an average age of $9.1 \pm 0.8$ Gyr and [$Z$/H]=$-1.2 \pm 0.2$ dex.

We find an observed scatter of 0.2 mag in $g-i$ and 0.3 in $u-i$. This scatter is consistent with the findings of \cite{Roman_21} (using DECam data) and \cite{Shen_21} (using 1-orbit \textit{HST} data), which also found a spread of 0.3--0.4 mag in the GC colours of DF2/DF4, but using different filter bands. The scatter is, however, broader than the recent findings of \cite{vanDokkum_22b}, who found that GCs in DF2/DF4 have an intrinsic spread in colours of less than 0.02 mag. This is not unexpected since we are using much shallower ground-based data. In contrast, \cite{vanDokkum_22b} applied a careful flat-fielding analysis on very deep \textit{HST} data (19 orbits for DF2 and 7 orbits for DF4) to get to their results, revealing that colour scatters can be found to be larger with lower quality data.

For consistency, using FSPS we converted our average $g-i$ colour of the GCs around DF2 and DF4 into $V_{606}-I_{814}$ to compare it with \cite{vanDokkum_22b}. We find that our colours translate to $V_{606}-I_{814} = 0.39 \pm 0.10$, which is consistent with their findings of $V_{606}-I_{814} = 0.375 \pm 0.005$. 
\cite{vanDokkum_22b} found the DF2 and DF4 GCs to be significantly bluer on average than GCs around low-luminosity Virgo cluster dEs \citep{Peng_06}. This conclusion, however, required a chain of three different conversions between $g-z$ and $V_{606}-I_{814}$ colours, which likely resulted in accumulated systematic errors. Here we apply a single conversion between $g_{475}-z_{850}$ and $g-i$ from \cite{Usher_12} (equation A1). We find that our average is $\langle g-z \rangle = 1.03 \pm 0.10$ and, thus, we find that the colours of the GCs around DF2/DF4 and around Virgo dwarfs \citep[$g-z \sim 0.9$ mag,][]{Peng_06} are consistent within the uncertainties.

\subsection{NGC~1052 GC density profile}
To study in detail the intragroup GC distribution and the GCs around the LSB galaxies, we have modelled and subtracted from our distribution the overpowering GC system of NGC~1052. For this analysis, we have used the broader GC selection based on the properties of GCs in general.

We first characterised the GC system of NGC~1052 by performing a radial number count analysis around the galaxy. In addition, the overall GC distribution around NGC~1052 was divided into blue ($0.4 < g-i \leq 0.85$) and red ($0.85 < g-i \leq 1.2$) GCs, following typical red/blue GC separations in the literature \citep[e.g.,][]{Lee_10}. We measured the average GC surface density around the galaxy in log spaced radial bins. 
Uncertainties were calculated by assuming Poisson statistics for the innermost bins. For the outer ones, in addition to the Poisson errors, we performed number count measurements with the same bin size in random places in our images to estimate the level of uncertainty in the background. These uncertainties were added quadratically.

We then fitted a S\'ersic function to the GC density profile of the overall GC distribution of NGC~1052, as well as the red and blue GC distributions. The S\'ersic index ($n$), GC half-number radius ($R_{\rm GC}$) and background+intragroup density ($\rho_N$ (bg+ig)) were free parameters. The profiles were fitted from the fourth radial bin on in order to avoid the incompleteness in the innermost bins due to the bright galaxy light.
The three resulting density profiles are shown in Fig. \ref{fig:densityprofile}.
The overall background and intragroup GC densities were calculated in isolated areas throughout the group, corresponding to $\rho_{\rm N(bg)} + \rho_{\rm N(intragroup)} = 0.27 \pm 0.03$ GCs/arcmin$^2$ and are shown in Fig. \ref{fig:densityprofile} as well. GCs are considered as not being bound to NGC~1052 once the GC number density reaches the background+intragroup level.
In Table \ref{tab:morphology_GCs}, we summarise the results of the model of these GC systems. We note that for all calculations of $R_{\rm GC}$/$R_{\rm e}$, we assumed the effective radius of the stellar body of NGC~1052 recovered by \protect\cite{Forbes_17}, i.e., $R_{\rm e} = 21.9$ arcsec.
We find that the overall GC distribution of NGC~1052 has a S\'ersic index of $3.04 \pm 0.02$ and $R_{\rm GC}$/$R_{\rm e}$ = $10.79 \pm 0.32$. The red GCs have a S\'ersic index of $2.92 \pm 0.12$ and $R_{\rm GC}$/$R_{\rm e}$ = $8.82 \pm 0.56$. The blue GCs have a S\'ersic index of $2.50 \pm 0.07$ and a more extended system with $R_{\rm GC}$/$R_{\rm e}$ = $13.20 \pm 0.43$. Additionally, as seen in Table \ref{tab:morphology_GCs}, the fit of the background density in the total GC profile is reassuringly consistent with the overall background density ($\rho_{\rm N(bg)} + \rho_{\rm N(intragroup)}$) measured in isolated areas, as previously mentioned.
The finding of blue GCs extending out to larger radii than red GCs is in agreement with the expectation that the metal-rich red GCs are mostly formed in-situ and have a shallower radial profile, while metal-poor blue GCs are mostly formed ex-situ and are later accreted onto the galaxy \citep{Brodie_06}. 

It is interesting to note that the GC distribution around NGC~1052 is quite extended for its stellar mass ($\log_{10}(M_{*}/M_{\odot}) = 11.02$, \citealt{Forbes_17}). \cite{Forbes_17} and \cite{Hudson_Robison_18} have shown that the sizes of the GC systems of galaxies scale with their stellar and halo masses. According to \cite{Forbes_17alone}, early-type galaxies (ETG) have an average $R_{\rm GC}/R_{\rm e}$ of $3.7 \pm 0.4$. \cite{Hudson_Robison_18}, similarly, found that the mean ratio is $R_{\rm GC}/R_{\rm e}$ = 3.5. However, both studies show that some galaxies can have much more extended GC systems, reaching in extreme cases even to a ratio of $R_{\rm GC}$/$R_{\rm e}$ = 46. We caution, none the less, that our recovered GC system sizes may be overestimated since the fits do not account for incompleteness in the central regions. This is not a problem for the works in the literature \citep{Forbes_17alone,Hudson_Robison_18} which applied methods of galaxy subtraction to better estimate the number of GCs in the innermost regions of the galaxies. A more detailed modelling of the central galaxy is beyond the scope of the paper.

After recovering the morphological parameters of the GC system, we reconstruct the GCLF of NGC~1052.
We assume a GCLF that is symmetric about the peak and that our distribution is complete down to the universal turnover magnitude (as discussed in Section \ref{sec:artstars}). We then double the number of GCs counted down to the turnover to estimate the total number of GCs around NGC~1052. 
Additionally, to correct for the missing innermost bins not included in the fit, we extrapolate the galaxy surface density profile to the centre, assuming the completeness calculated using our artificial star tests. Taking all of this into consideration, we find a final count of $402 \pm 9$ GCs around NGC 1052,
similar to the findings of \citealt{Forbes_01} (e.g., 390 GCs). 

The average colour of the GCs found around NGC~1052 is $\langle g-i \rangle = 0.84 \pm 0.16$ and $\langle u-i \rangle = 1.93 \pm 0.25$. These correspond to an average age of $11.2 \pm 1.6$ Gyr and [$Z$/H]=$-0.8 \pm 0.3$ dex, i.e., a significantly redder GC population than the one found around DF2/DF4 (see Section \ref{sec:DF2/DF4}).

\begin{figure}
    \centering
    \includegraphics[width=\columnwidth]{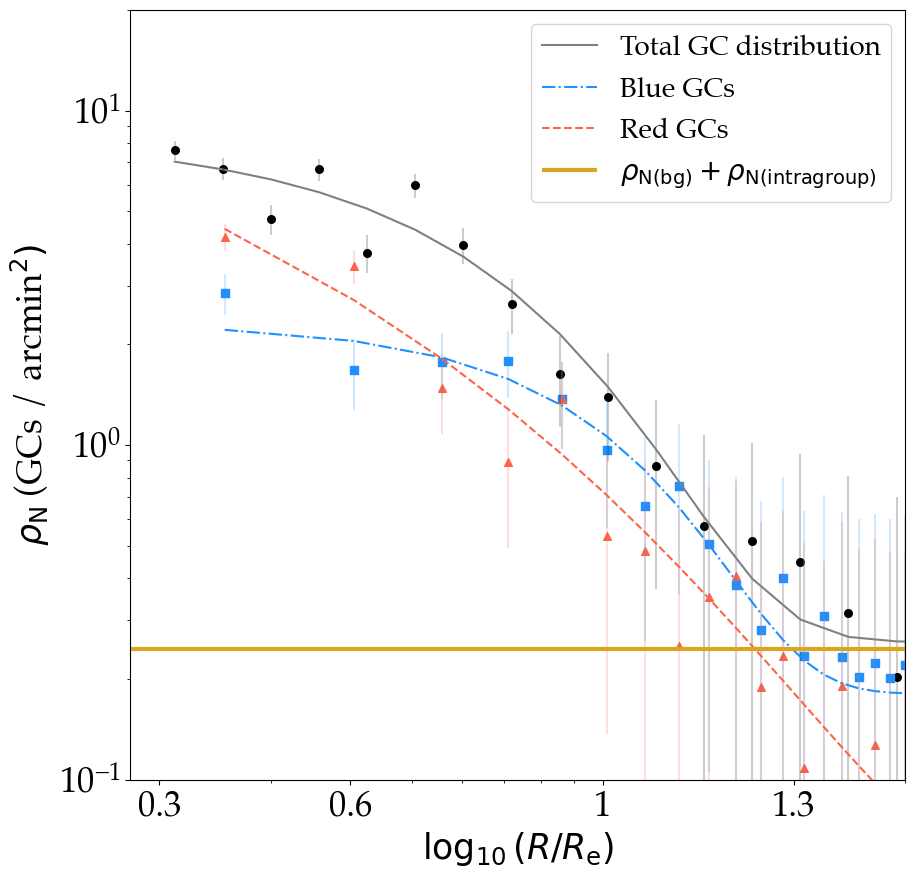}
    \caption{GC surface density profile around NGC~1052 from galaxy centre in terms of $R_{\rm e}$. Black circles, blue squares and red triangles stand for the total, blue and red GC number densities, respectively. The black, dashed red and dash-dotted blue curves are the S\'ersic model fits to the total, red and blue GC distributions, respectively. The solid yellow line is the GC number density contribution of the background and intragroup GCs measured in blank regions.}
    \label{fig:densityprofile}
\end{figure}

\begin{table*}
    \centering
    \begin{threeparttable}
    \caption{NGC 1052 GC system morphology.}
    \begin{tabular}{p{1.5cm}|p{2.0cm}|p{2.0cm}|p{2.0cm}}
         & \textbf{$n$} & \textbf{$R_{\rm GC}$/$R_{\rm e}$} & \textbf{$\rho_{\rm N(bg+ig)}$}   \\ \hline
    \textbf{All GCs}  & $3.04 \pm 0.02$ & $10.79 \pm 0.32$ & $0.25 \pm 0.05$ \\
    \textbf{Red GCs} & $2.92 \pm 0.12$ & $8.82 \pm 0.56$ & $0.06 \pm 0.04$ \\
    \textbf{Blue GCs} & $2.50 \pm 0.07$ & $13.20 \pm 0.43$ & $0.19 \pm 0.05$ \\ \hline
    \end{tabular}
    \begin{tablenotes}
      \small
      \item \textbf{Note.} Columns stand for: (1) Sources modelled; (2) S\'ersic index; (3) GC half-number radius relative to galaxy $R_{\rm e}$; (4) Background+intragroup GC number density in units of GCs/arcmin$^2$. These results were obtained using the GCs selected by the selection criteria based on SSPs. 
\end{tablenotes}
    \label{tab:morphology_GCs}
    \end{threeparttable}
\end{table*}

\subsection{Globular Cluster Large Scale Structure}

The 2D distribution of GCs around NGC~1052 was subtracted from the total GC distribution of the group using the recovered best fitting density profile, i.e., $R_{\rm GC} = 10.79\, R_{\rm e}$, assuming an axis ratio of 0.31 and position angle of 120 degrees measured by \cite{Forbes_01}. We assumed the same best fitting density profile to subtract the GCs around NGC~1052 from the map based on all GCs and the one based on the colours of DF2/DF4 GCs.
After subtracting the GC system of NGC~1052 itself, we can analyse the GC distribution around the fainter galaxies, as well as any intragroup GC structures. The recovered GC large scale structure based on SSP models is shown in Fig. \ref{fig:overall_map}. For comparison, we show the LSS based on the colours of the GCs around DF2 and DF4 in Fig. \ref{fig:narrow_map}.

By looking just at the GC large scale structure based on SSP models, we see a nearly uniform distribution, with the exception of GC overdensities around some of the giant galaxies in the group (e.g., NGC~1035 and NGC~1047), as well as many around NGC~1052 itself.
NGC~1052 was shown by \cite{Mueller_19} to have several LSB features around it, including stellar streams, arcs and tidal features. Specifically, \cite{Mueller_19} found a loop-shaped LSB feature to the south of NGC~1052 in the direction of NGC~1042 and another one forming an arc connecting NGC~1052 and NGC~1047. We highlight the approximate positioning of these features in our maps in Figs. \ref{fig:overall_map} and \ref{fig:narrow_map}, but see figure 1 of \citealt{Mueller_19} for further details. These features coincide with the positions of the GC overdensities found in our maps, and may be indicative of tidal interactions between the main galaxies in the group, with their outermost GCs being now part of these tidal features. In addition to \cite{Mueller_19}, \cite{vanGorkon_86} have also previously shown that the HI distribution of NGC~1052 has tidal features extending in the same direction as these identified GC overdensities, once again indicating that the galaxies may be interacting \citep[see figure 2 of][]{Mueller_19}. 

The GC map based on DF2 and DF4 selection has very few overdensities since it is based on very narrow selection criteria. Reassuringly, we recover the overdensities around DF2 and DF4 and very weak overdensities around RCP32 and DF9. We highlight in both maps the positions of LSB galaxies found by \cite{Roman_21}. We investigate if there are any GCs around them or intragroup GC overdensities connecting them. No clear overdensities were found throughout the group, only around NGC~1052 itself.
Again, these share the same positions as previously identified tidal tails and stellar streams around NGC~1052 \citep{Mueller_19}. 

In the top row of Fig. \ref{fig:overall_map}, we show the four dwarf galaxies identified to have GC systems: RCP 32 ($N_{\rm GC} = 2$), DF2 ($N_{\rm GC} = 11$), DF9 ($N_{\rm GC} = 3$) and DF4 ($N_{\rm GC} = 10$), respectively. We note that each stamp has 1 arcmin$^2$, and the background+intragroup contamination was found to be $\rho_{\rm N (bg+ig)} = 0.27$ GCs/arcmin$^2$.
These findings are discussed in detail in the following section.

\begin{figure*}
    \centering    
    \includegraphics[width=0.515\columnwidth]{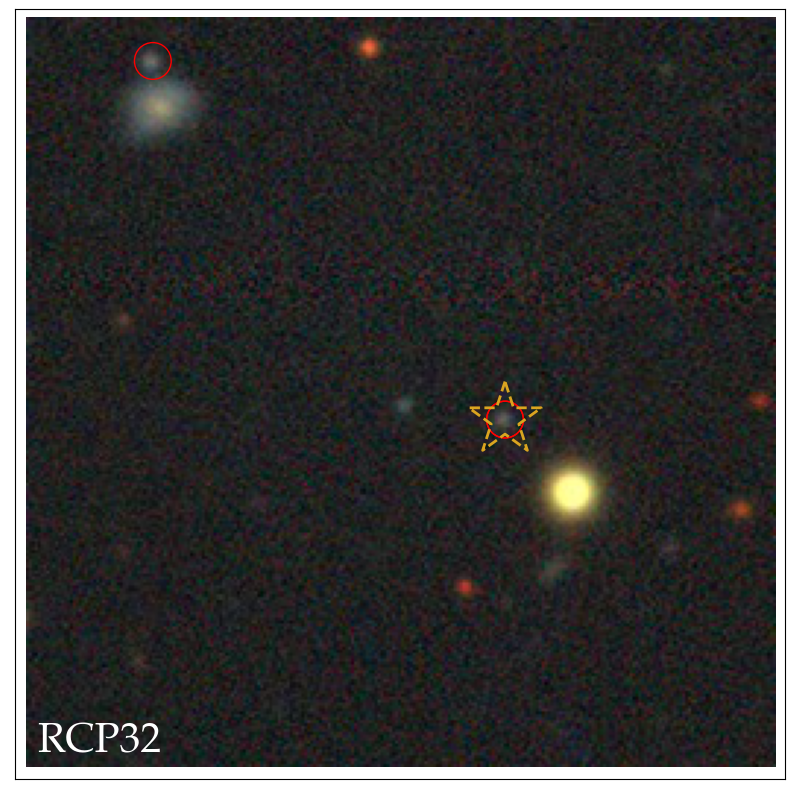}
    \includegraphics[width=0.515\columnwidth]{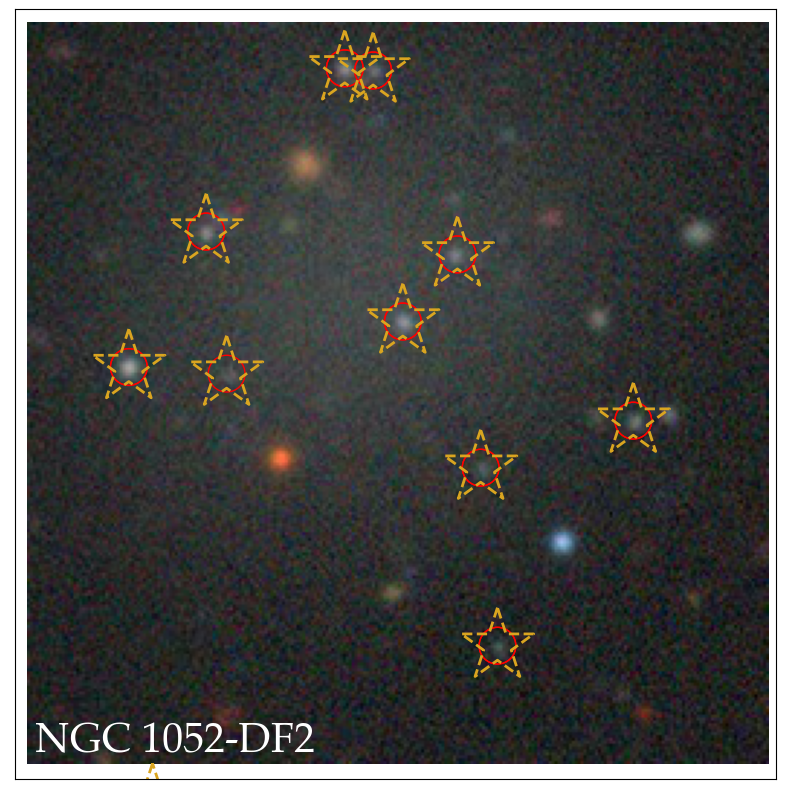}
    \includegraphics[width=0.515\columnwidth]{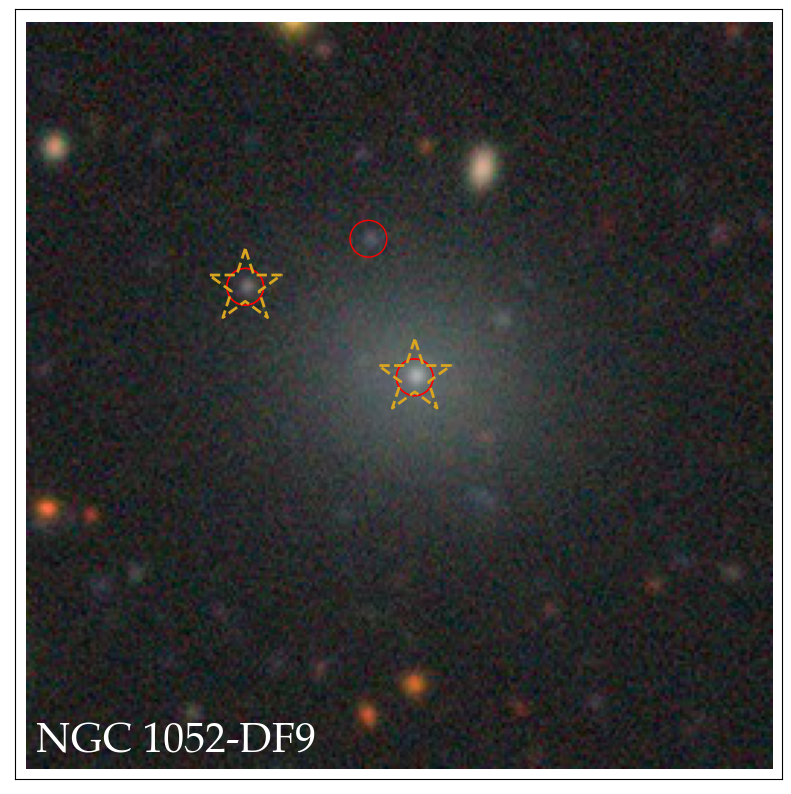}
    \includegraphics[width=0.515\columnwidth]{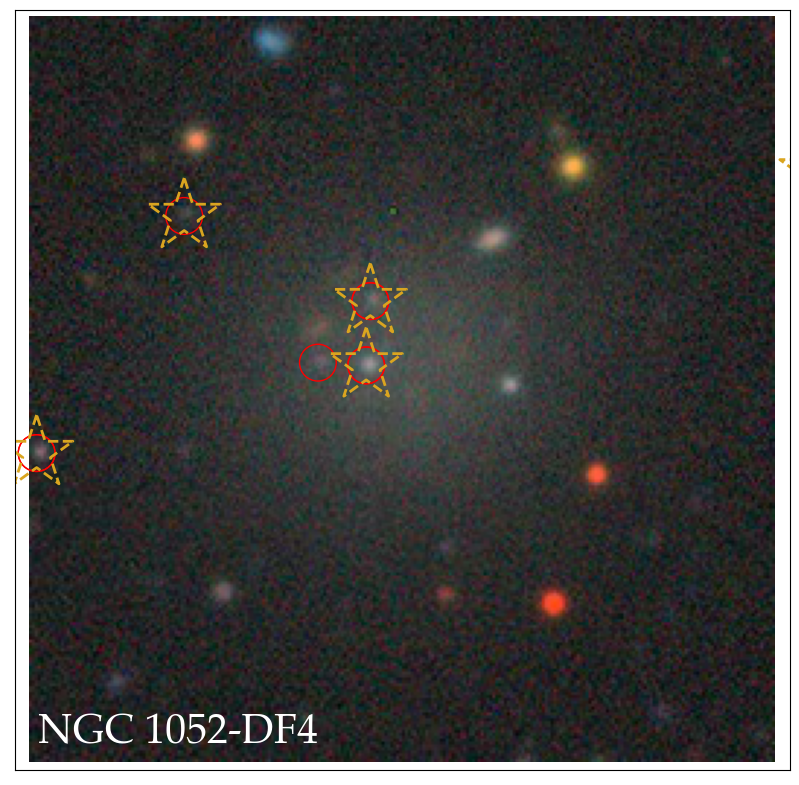}
    \includegraphics[width=\textwidth]{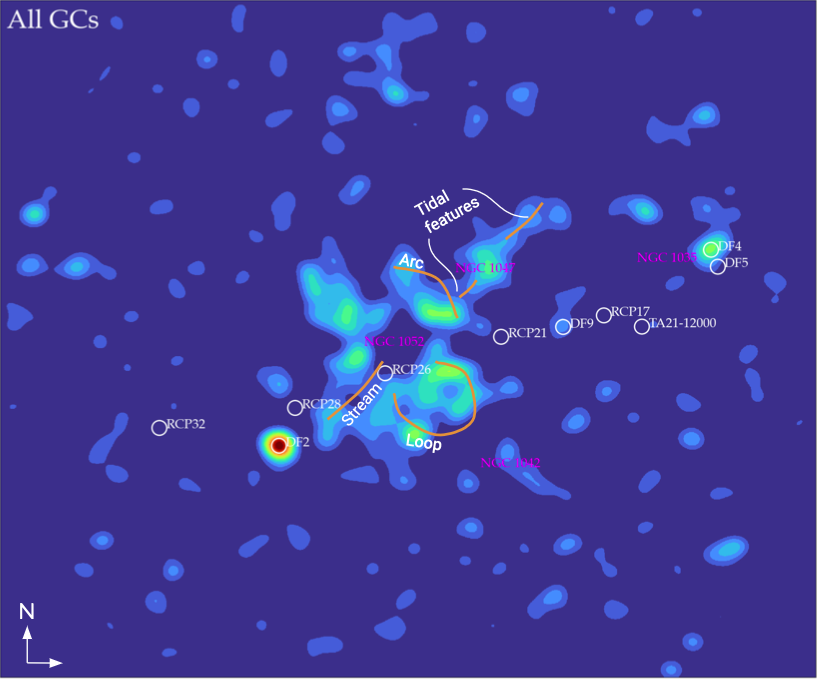}
    \caption{NGC~1052 group globular cluster large scale structure.
    Top row: $g$ band stamps of four dwarf galaxies in the group (RCP32, DF2, DF9 and DF4), respectively. North is up, east is left. Each stamp is 1 arcmin on each side. The yellow stars are the GCs selected according to the properties of the spectroscopically confirmed GCs around DF2/DF4. Overlaid as red circles are the GCs selected using the colour selection based on SSP models. Bottom: GC large scale structure based on SSP models. The figure has a size of 1.2 x 1 sq. degrees. 
    The structure has had the GC system of NGC~1052 subtracted. White circles show the LSB galaxies identified by \protect\cite{Roman_21}. The magenta labels are all of the large galaxies in the group. The orange curves show the previously identified stellar LSB features in the group \protect\citep{Mueller_19}. 
    The map reveals GC overdensities around NGC~1052 itself, in the same positions as previously identified stellar streams in the group.}
    \label{fig:overall_map}
\end{figure*}

\section{Discussion}
\label{sec:discussion}

The 2D distribution and colours of GCs across the group and on its galaxies can tell us much about the physical processes taking place in that environment. \cite{Lee_10}, for example, found many blue intracluster GCs forming tidal features around the bright elliptical galaxies in the Virgo cluster. The distribution of these intracluster GCs and the fact that they were blue led them to believe that these were stripped from low-mass dwarfs and were now being accreted to nearby massive galaxies/to the centre of the cluster. Although applied to a larger scale, the study of \cite{Lee_10} highlights that the intracluster/intragroup GC distribution is crucial to differentiate formation scenarios and processes undergoing in group and cluster environments. The existence of intragroup/intracluster GCs can be explained by gas collapse outside of the galaxies or by stripping GCs initially bound to galaxies. The distribution and stellar populations of intragroup GCs, thus, can be indicative of the way they were formed.

In the following sections, we analyse different formation scenarios for the galaxies in the NGC~1052 group. We will look particularly at their feasibility to form the GCs observed in the galaxies and in the intragroup medium, i.e., if they are able to reproduce the recovered GC LSS.

\begin{figure}
    \centering    
    \includegraphics[width=\columnwidth]{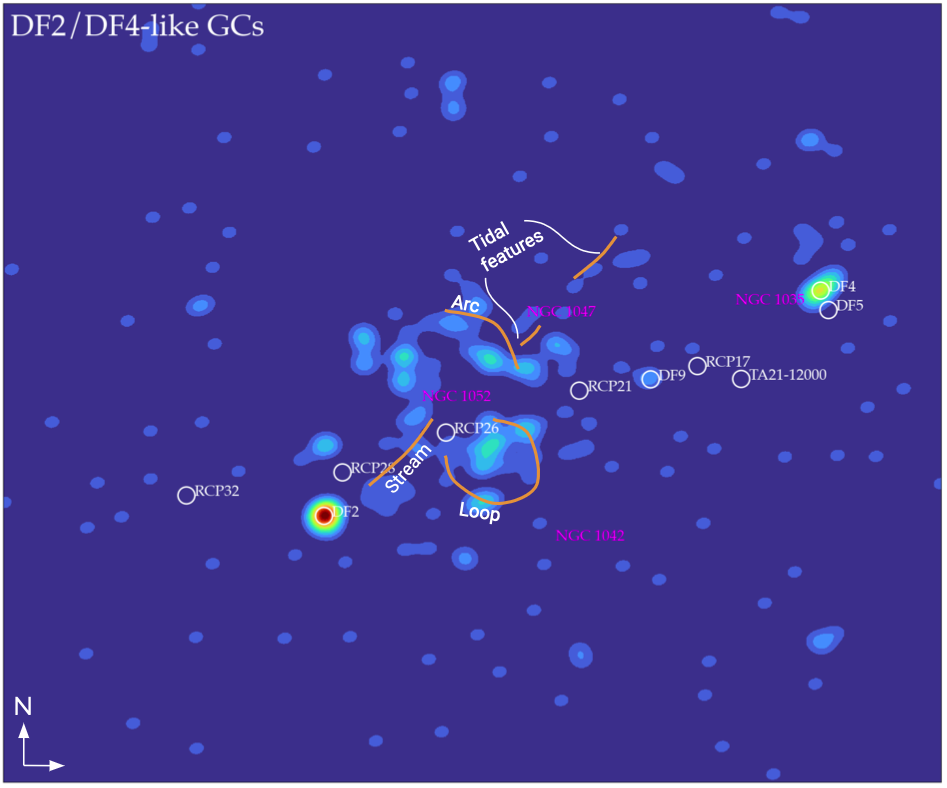}
    \caption{GC large scale structure based on the properties of spectroscopically confirmed GCs around DF2/DF4. Map has a size of 1.2 x 1 sq. degrees. The structure have had the GC system of NGC~1052 subtracted. White circles show the ten LSB galaxies present in the `bullet trail' (DF7 is out of our FoV). The magenta labels are all of the large galaxies in the group. The orange curves show the previously identified stellar LSB features in the group \protect\citep{Mueller_19}. The figure shows that there is no significant GC overdensity along the trail with the same colours as DF2/DF4 GCs. The map also reveals the same GC overdensities around NGC~1052 seen in the map based on SSP colours.}
    \label{fig:narrow_map}
\end{figure}

\subsection{Bullet-dwarf formation}

The bullet-dwarf scenario makes distinct predictions for the stellar populations (and thus colours) of all of the supposedly DM-free galaxies along the trail, as well as for their GC systems. In this Section, we explore how well the GC distribution around the dwarf galaxies in the group agree with this formation scenario.
Additionally, we investigate the presence of intragroup GCs that would have been formed by the same process. If they do exist, they would be expected to have similar stellar populations to the GCs around DF2 and DF4, and also, due to conservation of momentum and inertia, these GCs would be expected to follow the same near-linear spatial distribution as the trail galaxies. The schematic picture, shown in figure 1 (appendix) of \cite{vanDokkum_22}, suggests that some GCs may be formed away from DF2/DF4, i.e., in other trail galaxies and/or as intragroup GCs along the trail.

To be consistent with the scenario, all of the GCs are expected to have an age consistent with $\sim$8 Gyr, as found for the stellar body of DF2 and DF4 by \cite{Buzzo_22b}. This finding is supported by \cite{Fensch_19}, which using IFU (MUSE/VLT) data, have shown that the stellar body and GC population of DF2 have equivalent ages (8.9 Gyr). Additionally, \cite{vanDokkum_22b} have shown that the GC systems of DF2 and DF4 have the same optical colour, thus, same stellar populations. All of these seem to support the idea that DF2, DF4 and their GCs had a coeval formation. 

In what follows, we investigate the other dwarfs in the group found to host GC systems. Combining these findings with what we know about DF2 and DF4, we discuss the level of agreement of the formation scenario with the recovered LSS.

\subsubsection{RCP32}

We find two GCs around RCP32, one of the expected DM-dominated galaxies in the trail. We cross-matched our GC findings around RCP32 with those found by \cite{Roman_21}, resulting in four GCs in common out of a total of eight GCs identified by them. With visual inspection, we confirmed that the additional four GC candidates identified by them are extended in our imaging (i.e., likely background galaxies). Out of the four GCs in common, we conclude that only two of them are likely part of RCP32, since the other two are too distant from the galaxy. One of the two GCs found to be close to RCP32 has similar colours to those of DF2 and DF4 and it is luminous with $M_g = -8.2$ ($g = 23.5$) mag. The other GC is bluer ($u-i = 1.52 \pm 0.08$, $g-i = 0.79 \pm 0.07$) than those around DF2 and DF4, but it meets the colour criterion based on DF2/DF4 within the uncertainties. The average colours of the GCs around RCP32 are $\langle g-i \rangle = 0.82 \pm 0.01$ and $\langle u-i \rangle = 1.58 \pm 0.06$. These correspond to an age of $\sim 11.2 \pm 0.7$ Gyr and [$Z$/H]=$-1.4 \pm 0.2$ dex. The age of these GCs, thus, is more similar to the ones found around NGC~1052 than the ones around DF2/DF4.

This galaxy, in fact, is not expected to host GCs formed in the interaction, but rather to host the stars and GCs of the progenitor galaxies (see extended data figure 1 of \citealt{vanDokkum_22} for further details). This is due to the collisionless nature of dark matter, stars and primordial GCs, which simply pass each other in the interaction. The geometry of the collision, conservation of momentum and inertia cause them to be located at the ends of the trail. In contrast, all of the new galaxies and GCs formed from the collisional gas will be distributed along the trail. Thus, finding a GC with different colours from those of DF2 and DF4 is not evidence against a bullet-dwarf formation in this case, but indicative that this GC may have come from one of the progenitor galaxies.

\subsubsection{DF9}

We find three GC-like sources around DF9. It is interesting to note that the object identified in the centre of DF9 is likely the nucleus of the galaxy, and not actually a GC. However, because of its overluminous nature ($M_g = -9.6$ or $g = 21.9$ mag), this object was selected for having properties similar to those of the overluminous GCs around DF2 and DF4. 

The nucleus was found to have a colour of $g-i = 0.71 \pm 0.06$ and $u-i = 1.68 \pm 0.07$, which, comparing to SSP models, indicates an age of $9.4 \pm 1.1$ Gyr and [$Z$/H] = $- 1.2 \pm 0.2$ dex. This is consistent with the ages and metallicities of both the stellar body and nucleus of DF9 obtained with Keck/KCWI spectroscopy (Gannon et al. in prep.), and in line with the stellar populations of the GCs in DF2/DF4 \citep{Fensch_19}.

Apart from the nucleus, we are left with only two GCs around DF9. As found for RCP32, one of the GCs has similar colours and brightness to those of DF2 and DF4 and the other is bluer ($g-i = 0.68 \pm 0.05$, $u-i = 1.45 \pm 0.12$), but again consistent with the properties of DF2/DF4 within the uncertainties. The average colours of the GCs around DF9 are $\langle g-i \rangle = 0.75 \pm 0.06$ and $\langle u-i \rangle = 1.63 \pm 0.20$. These correspond to an age of $\sim 8.2 \pm 0.9$ Gyr and [$Z$/H]=$-1.2 \pm 0.2$ dex, which are consistent with the stellar populations of the GCs around DF2/DF4.

\subsubsection{Agreement with the hypothesis?}

The bullet-dwarf collision scenario makes the prediction that any GCs formed in the interaction should have small variations in colour and all be consistent with an age of $\sim$8 Gyr.
The model suggests that the metallicities of the sources should be similar, but does not predict any specific value as it does for the ages. Previous studies, none the less, have shown that DF2 and DF4 have stellar metallicities of [$Z$/H] $\sim -1.1$ dex \citep{Fensch_19,Buzzo_22b}. If the other galaxies/GCs were formed from the same material, there is an expectation that they would have the similar metallicities. The GC colours and the stellar body of DF2 and DF4 were shown to meet these expectations \citep{Buzzo_22b,vanDokkum_22b}.
Apart from DF2 and DF4, only RCP32 and DF9 are found to host GC populations (both with $N_{\rm GC} \leq 3$). No intragroup structure was identified along the bullet trail, which naturally raises the question of why the collision would form a trail of galaxies, but only a portion of them would have enough gas to form several and mostly overluminous GCs. 

The finding of two GCs around RCP32 (one of the DM-dominated dwarfs) with colours similar to those of the progenitor galaxy (NGC~1052) and three GCs around DF9 with equivalent colours  (within the uncertainties) to those of DF2 and DF4 GCs are consistent with the predictions of the scenario. 

To further investigate this finding, we estimate, based on the total luminosity of the trail, the predicted number of GCs in the trail detectable within our imaging limitations. 
Under the assumption that the trail was formed according to a bullet-dwarf scenario, the galaxies would have been formed by the same gas with the same GC formation efficiency. Thus, it is a fair assumption that all galaxies should have similar GC specific frequencies \citep[$S_N$,][]{Harris_81}, which can be defined as:

\begin{equation}
    S_N = N_{\rm GC} \times 10^{0.4 (M_V + 15)}
\end{equation}
 
With this in mind, we assume that all of the galaxies share the same specific frequency as the combined $S_N$ of DF2 and DF4 --the trail galaxies with the most GCs per unit luminosity ($S_N = 14.6$). 
Although this may be a strong assumption (i.e., the same GC $S_N$ for all galaxies), it puts at least a rough estimate on the number of GCs we can expect throughout the trail and if GC formation proceeded as in DF2 and DF4.
We note the caveat that $S_N$ is dependent on the number of GCs and, consequently, dependent on the imaging depth, i.e., how many GCs can be detected within our imaging limitations. For example, the $S_N$ quoted in this study is smaller than the one found by \cite{vanDokkum_18,vanDokkum_19} for DF2 (i.e., $S_N \approx 11$, \citealt{vanDokkum_18}) and DF4, both recovered using deeper HST data. Thus, we caution that our estimates of the number of GCs in other trail galaxies are biased by what is detectable in our imaging.

Using this information and the luminosity of each trail galaxy from \cite{Roman_21}, we find that the predicted number of GCs for the whole trail is $N_{\rm GC} = 42.3 \pm 4.7$. 
As expected, most of these GCs are located around DF2 and DF4. The galaxy with the next highest number of predicted GCs is DF9, with $N_{\rm GC} = 3.6 \pm 1.2$, which is consistent with our finding of 3 clusters. All of the other galaxies were found to have a predicted number of GCs smaller than 2, with RCP32 having $N_{\rm GC} = 0.02 \pm 0.02$. These are only approximate values, but they confirm expectations of a very small number of GCs around the trail galaxies, apart from DF2 and DF4. Reassuringly, the expected and observed number of GCs around DF9 are consistent. On the other hand, although the expected number of GCs around RCP32 may be smaller than observed, we reemphasise that the GCs around RCP32 are likely from the pre-collision galaxy. Thus, the assumption that this galaxy has formed GCs with the same efficiency as DF2/DF4 is not expected to hold. 

Thus, some of our findings seem to be broadly consistent with the bullet collision, namely: 1) The GC system around RCP32 has stellar populations similar to those of the progenitor galaxy. 2) The GC system around DF9 has populations similar to those of the GCs around DF2 and DF4. 3) The predicted number of GCs around DF9 (3.6) is consistent with the finding of 3 GCs around it. 4) The lack of GCs around other dwarfs is consistent with the prediction that their GC numbers are smaller than 1. While these aspects may be broadly consistent with the scenario, the lack of constraints on the formation histories of any of the other galaxies in the trail prevents us from conclusively adhering to this scenario. 
The best way forward is to study the stellar populations of the galaxies themselves to further point to this scenario as a possibility (Tang et al. in prep.).

\subsection{Tidal Dwarf Galaxies}

In a group with so many LSB features \citep{Mueller_19}, the existence of the NGC~1052 dwarfs could be the result of galaxy-galaxy interactions, leading to tidal dwarf galaxies \citep[TDGs,][]{Haslbauer_19, Mueller_19}. 
This particular mode of formation typically leads to young, metal-rich, gas-dominated (unlike DF2 and DF4) and dark-matter free galaxies. Simulations have shown, none the less, that it is possible for TDGs to be formed at high-redshifts ($0.5 < z < 2.0$), resulting in old and quiescent dwarfs observed at $z=0$ \citep{Ploeckinger_18}. This scenario for the formation of the NGC~1052 dwarfs would be similar to the idea proposed for the Milky Way plane of satellites \citep{Lynden-Bell_76} or to that of the Dentist's Chair \citep{Weilbacher_02}. Additionally, if the galaxies in the NGC~1052 group were formed by this scenario, they would have formed from gas and stars originating from the outskirts of a massive host after a galaxy-galaxy interaction. In this case, we may expect intragroup GCs with similar stellar populations as the ones from the progenitor massive galaxy, as they would have been stripped together with the gas that formed the TDGs. 

This formation scenario successfully explains the DM depletion, and it may also explain the normal GC population of the dwarf galaxies. However, this scenario, as it stands, cannot explain the overluminous population of GCs around DF2 and DF4. This is because low-redshift TDGs do not have star formation rates high enough to form overmassive GCs \citep[][and reference therein]{Fensch_19b}.
In fact, \cite{Fensch_19b} suggest that although TDGs created by major mergers can lead to the formation of massive young clusters, their survival times are expected to be only of some hundreds of Myr. None the less, they hypothesise that if similar TDGs would have been formed at high-redshift ($0.5 < z < 2.0$), where gas fractions were higher and could lead to stronger star formation episodes, then the formation of ultramassive GCs that would survive for a Hubble time is possible.

Another problem encountered by the tidal dwarf scenario for DF2 and DF4 is that the gas content forming TDGs and their GCs is supposed to be pre-enriched in metals, leading to metallicities of one third to half solar. Thus, they are expected to have significantly higher metallicities than those observed for the GCs in this study, i.e., average [Z/H]$ = -1.2 \pm 0.2$ dex. However, if we once again resort to high-redshift TDGs \citep{Fensch_19b}, these would have been formed at a time where the interstellar medium would not have been as pre-enriched and, thus, their GCs would possibly display metallicities closer to the observed values. The formation of TDGs and their GCs at high-redshift ($z \sim 1.9$) would also be in agreement with the average age of the GCs in the dwarf galaxies of $9.1\pm 0.9$ Gyr.

High-redshift TDGs, thus, seem to be able to explain many of the properties of the dwarfs in the NGC~1052 group, i.e., DM depletion, overluminous GC population and low metallicities.
One problem though is that the two DM-free dwarf galaxies in the NGC~1052 group were shown to be close only in projected distance, but far apart in the line-of-sight TRGB distance \citep{Shen_21}, which is hard to reconcile with any tidal dwarf-like formation at low- or high-redshift, as TDGs by definition never leave the gravitational potential of their progenitors, being always within one or two hundred kpc from them \citep{Ploeckinger_18}.

We conclude that a high-redshift tidal dwarf galaxy model as it is, similarly to the bullet-dwarf scenario, has points of agreement with our findings, e.g., GC metallicities and ages; and aspects that need further investigation, i.e., physical separation of dwarf galaxies.
Simulations of high-redshift TDGs that can overcome the physical separation inconsistencies would provide a consistent formation scenario for the galaxies and GCs in the NGC~1052 group.

\subsection{Tidal Stripping}
Since the first claim that the galaxies in the NGC~1052 group could be dark matter-free, the most prominent model to explain their formation relied on tidal stripping by a more massive galaxy \citep[e.g., ][]{Ogiya_18, Ogiya_21, Moreno_22}. Such models were shown to successfully explain the lack of DM and the presence of GCs in the galaxies, but even using different simulations, they were not able to create GCs as massive as the ones seen in DF2 and DF4. 

Tidal stripping models, additionally, are expected to leave a trail of stripped GCs between the stripped dwarf galaxy and the massive host \citep{Lee_10}, but such structures are not seen in any of our LSS maps.

On the other hand, GC streams and other GC LSS structures are observed all around NGC~1052, indicating recent interactions between NGC~1052, NGC~1047 and NGC~1402, as suggested by \cite{Mueller_19}. Thus, tidal stripping could explain the connections found between the massive galaxies in the group, but do not seem to have any connection with the dwarfs.

\subsection{Modified Gravity}
Models that are alternative to the existence of dark matter may also be suitable explanations for the existence of these DM-free galaxies \citep[e.g., MOND; ][]{Kroupa_12}. They have been proposed as possible formation pathways for DF2 by \cite{Kroupa_18} and for DF4 by \cite{Mueller_19b}. These models, however, do not make predictions about the globular cluster population of the galaxies nor about their stellar populations, thus, they are more difficult to test with this study. They remain a possibility to be further developed and tested.

\vspace{-0.4cm}
\section{Conclusions}
In this work, we investigated the large scale structure of GCs in the NGC~1052 group in order to test the level of agreement of different formation scenarios proposed for the dwarf galaxies in the group with the recovered GC LSS. These include the `bullet-dwarf' scenario, high-redshift tidal dwarf galaxies, tidal stripping and modified gravity models. Using CFHT data in the $u$, $g$ and $i$ bands, we select two sets of GC candidates: one based on SSP models consistent with GCs in general, with a typical range in age of $8 - 14$ Gyr and metallicity of $-2.2 <$ [$Z$/H] $< - 0.0$ dex; and a second narrower one based on the properties of spectroscopically confirmed GCs around DF2 and DF4.
We subtracted the GC system of NGC~1052 to be able to see smaller overdensities around all LSB galaxies and any intragroup GC structures. The distribution of GCs around NGC~1052 was modelled to be larger than normal for an early-type galaxy of its mass, reaching a GC number radius of 10.8 galaxy effective radii. If the GC system of NGC~1052 is separated into red and blue GCs, these extend out to 8.8 and 13.2 $R_{\rm e}$, respectively. The GCs around NGC~1052 have an average colour that correspond to an age of $11.2 \pm 1.6$ Gyr and [$Z$/H]$=-0.8 \pm 0.3$ dex.

The GC map based on SSP colours revealed overdensities in the centre of the group, around NGC~1052 itself, and GC systems around the giant galaxies in the group. The overdensities around NGC~1052 coincide with the positions of pre-identified stellar streams and LSB features in the group, indicating past tidal interactions.
The GC map based on the properties of spectroscopically confirmed GCs around DF2 and DF4 also revealed overdensities in the centre of the group, in the same positions as the previously identified stellar streams. Reassuringly, the GC systems of DF2 and DF4 themselves were also found. They have an average age of $9.1 \pm 0.8$ Gyr and [$Z$/H]$=-1.2 \pm 0.2$ dex, being thus younger and less metal-enriched than NGC~1052.
No prominent intragroup GC structures were found in any map. On the other hand, GC systems were identified around four dwarf galaxies in the group: DF2 and DF4, as already known, but also RCP32 with 2 GCs and DF9 with 3 GCs. The average age of the GCs around RCP32 is consistent with the age of the GCs around NGC~1052, while the GCs around DF9 are consistent, within the uncertainties, with the colours of DF2/DF4 GCs.

If we assume a bullet-like formation and that all of the galaxies in the supposed trail share the averaged specific frequency of DF2 and DF4 --the trail galaxies with most GCs per unit luminosity ($S_N$=14.6)-- we predict a total number of GCs throughout the trail of $N_{\rm GC} = 42.3 \pm 4.7$. Our predictions indicate that the number of GCs around DF9 is $N_{\rm GC} = 3.6 \pm 1.2$, which is consistent with our finding of 3 GCs. All of the other galaxies in the trail were found to have a combined predicted number of GCs smaller than 2. These predictions are consistent with the lack of GCs found in any of the other trail galaxies or in the intragroup medium. 
In addition, the colours (thus, stellar populations) of the GCs found around DF9 are similar to the ones around DF2 and DF4 and, thus, meet the stellar population expectations of the scenario. On the other hand, the GCs around RCP32 have equivalent ages to the GCs around NGC~1052 itself, which is consistent with the expectation that this DM-dominated dwarf would host the primordial GCs coming from the progenitor galaxy. While this scenario has many aspects of agreement with the recovered GC LSS, the lack of conclusive evidence on the properties of any of the other trail galaxies makes it hard to reach any definite conclusions on their formation scenario.

We discuss how TDGs formed at high-redshifts ($0.5 < z < 2$), when merger and star formation rates were higher, may be able to explain the overluminous population of GCs found in the galaxies. As they would have been formed at high-$z$, the metallicities and ages of the GCs may be consistent with the ones observed in this study.
TDG models, however, at low or high-redshift, do not seem to be able to explain the 2 Mpc distance between the two DM-free dwarf galaxies in the group, as this mode of formation predicts that the galaxies will live within about one or two hundred kpc from their progenitors. Thus, although promising, models of high-redshift TDGs need further exploration in order to correctly explain the physical separation of the galaxies. 

We discuss how tidal stripping models are unlikely to explain the dwarf galaxies in the group, as they cannot explain the overluminous GCs in the galaxies. We suggest, however, that tidal stripping could be taking place around the massive galaxies in the group, such as NGC~1052, NGC~1047 and NGC~1042, as stellar streams and tidal features are observed connecting these galaxies and are key imprints of this type of interaction. Additionally, we argue that modified gravity models remain as a possibility to explain these DM-free galaxies, but as they do not make any predictions on the stellar populations or GC systems of the galaxies, they are more difficult to test with our study.

We conclude that the GC distribution in the group does not conclusively point to any formation scenario in particular, but it favours models relying on galaxy-galaxy interactions and on the coeval formation of dwarfs and their GC systems. These include the recently proposed bullet-dwarf formation, as well as tidal dwarf galaxy models. Both scenarios need further exploration and adjustments to be fully compatible with the GC LSS observed in the group.

\label{sec:conclusions}

\vspace{-0.4cm}
\section*{Acknowledgements}
We thank the referee for several useful comments.
This research was supported by the Australian
Research Council Centre of Excellence for All Sky Astrophysics in 3 Dimensions (ASTRO 3D), through project number CE170100013. We thank the ARC for financial support via DP220101863. AJR was supported as a Research Corporation for Science Advancement Cottrell Scholar. 

This study made use of archival data and meta-data provided by the CFHT Science Archive (CADC). It was based on observations obtained with MegaPrime/MegaCam, a joint project of CFHT and CEA/DAPNIA, at the Canada-France-Hawaii Telescope (CFHT), which is operated by the National Research Council (NRC) of Canada, the Institut National des Science de l'Univers of the Centre National de la Recherche Scientifique (CNRS) of France, and the University of Hawaii. 
The observations at the Canada-France-Hawaii Telescope were performed with care and respect from the summit of Mauna Kea which is a significant cultural and historic site.

\section*{Data Availability}
The CFHT data are available from the \href{https://www.cadc-ccda.hia-iha.nrc-cnrc.gc.ca/en/search/?collection=CFHT&noexec=true}{CADC}. 

\vspace{-0.4cm}



\bibliographystyle{mnras}
\bibliography{bibli} 




\appendix




\bsp	
\label{lastpage}
\end{document}